\documentclass[a4paper]{jpconf}
\usepackage[dvips]{graphicx}
\usepackage{lineno}
\usepackage{textcomp}
\def \pao {~[Pierre Auger Collaboration]}
\def \fullauthor {Pierre Auger Collaboration [J. Abraham et al.]}
\def \icrcmx {(ICRC 2007)}
\def \icrcin {(ICRC 2005)}
\def \icrchb {(ICRC 2001)}
\begin{document}
\title{Atmospheric Calorimetry above 10$^{19}$ eV: Shooting Lasers at the Pierre Auger Cosmic-Ray Observatory}

\author{Lawrence Wiencke for the Pierre Auger Collaboration}

\address{Department of Physics, Colorado School of Mines, 1532 Illinois St, Golden CO, 80401 USA}

\ead{lwiencke@mines.edu}

\begin{abstract}
The Pierre Auger Cosmic-Ray Observatory uses the earth's atmosphere as a calorimeter
to measure extensive air-showers created by particles of astrophysical origin.  Some
of these particles carry joules of energy. At these extreme energies, test beams are
not available in the conventional sense. Yet understanding the energy response of the
observatory is important. For example, the propagation distance of the highest energy cosmic-rays through the cosmic microwave background radiation (CMBR) is predicted to be strong function of energy.  This paper will discuss recently reported results from the
observatory and the use of calibrated pulsed UV laser "test-beams" that simulate the optical signatures of ultra-high energy cosmic rays. The status of the much larger 200,000 km$^3$ companion detector planned for the northern hemisphere will also be outlined.

\end{abstract}

\section{Introduction}
\begin{figure}[h]
\begin{minipage}{16pc}
\includegraphics[width=16pc]{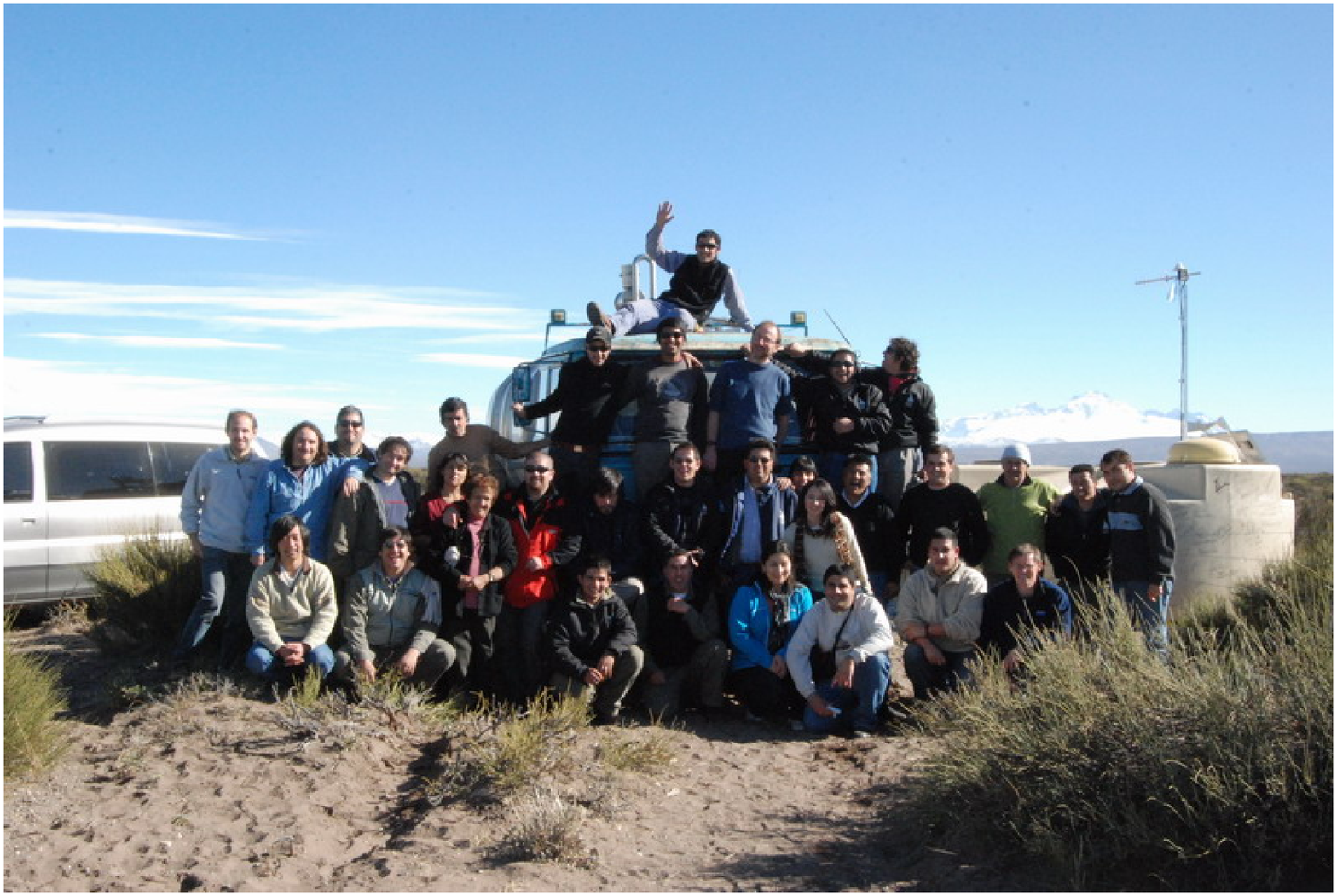}
\caption{\label{tlast}June 13th 2008: Auger South staff celebrate deployment of the commemorative final surface detector tank. The design specification of 1600 operation tanks has been achieved.}
\end{minipage}\hspace{2pc}%
\begin{minipage}{22pc}
\includegraphics[width=22pc]{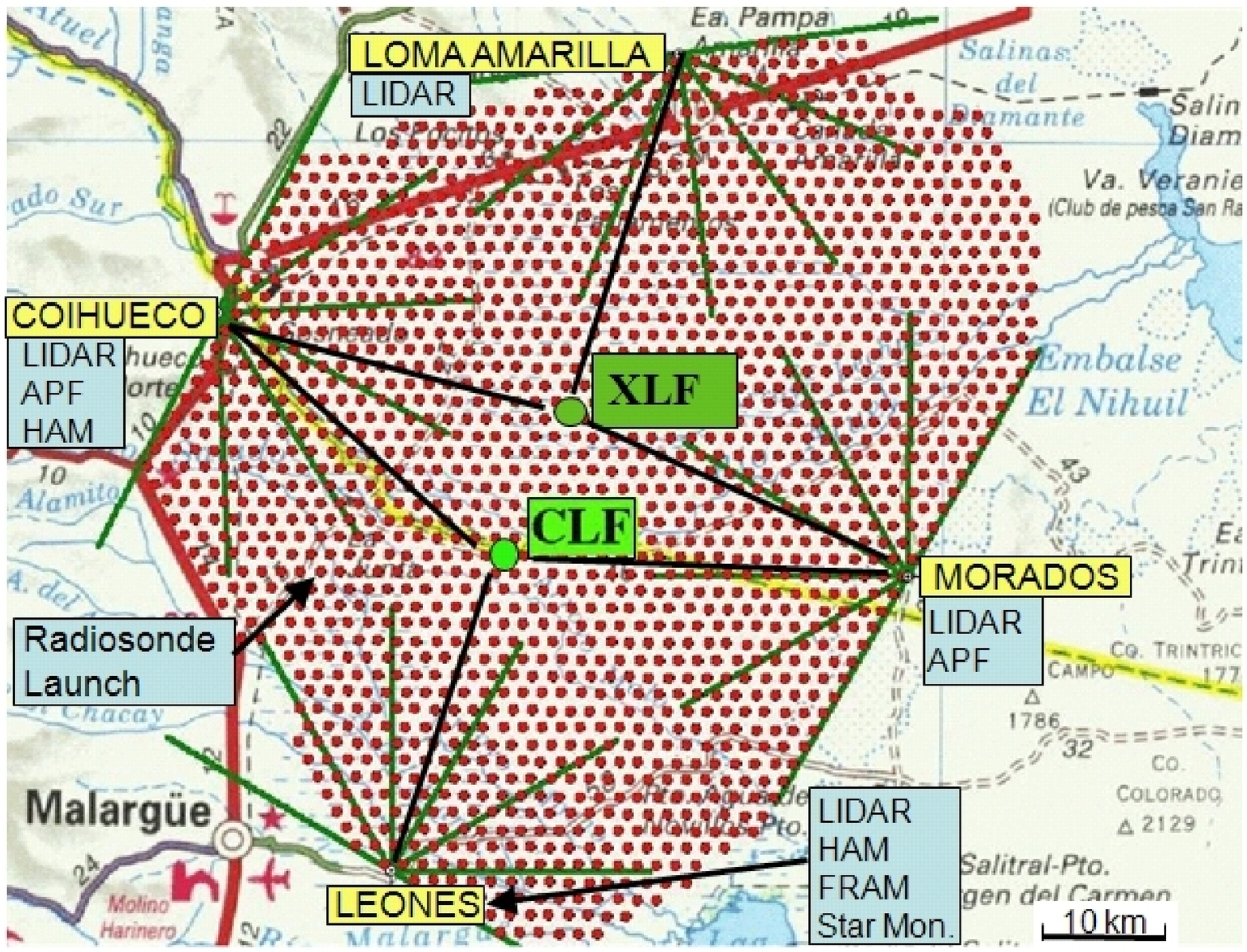}
\caption{\label{config}Auger South baseline configuration.}
\end{minipage} 
\end{figure}

By the time this article appears in print, the base-line configuration of Pierre Auger Detector in Argentina (Auger South) will be completed. The tank commemorating completion of the surface detector was activated (Fig. \ref{tlast}), two weeks after this conference concluded.

Roughly the size of Provincia di Pavia, Auger South is the largest detector of cosmic rays above $5\times10^{18}$eV in the world. The instrument has been described in previous CALOR conferences \cite{bianca06} and elsewhere \cite{paoprotonim}. So this talk will focus instead on an aspect of this experiment that has not been presented in this conference series yet relates closely to its calorimetric theme. This is the use of atmospheric laser \textgravedbl test-beams\textacutedbl \space that simulate many of the optical properties of an extensive air-shower (EAS).

\section{The Pierre Auger Southern Detector}
A co-located surface detector (SD) and air-fluorescence detector (FD) comprise Auger South. The SD includes an array of more than 1600 water cherenkov detector stations \cite{sdtank} on a 1.5 km grid, 4 fluorescence \textgravedbl eyes\textacutedbl, two test-beam laser facilities and an extensive collection of instrumentation (\cite{fram}, \cite{apf}, \cite{baloons}, \cite{lidar}) dedicated to atmospheric monitoring (\cite{aeromea07}, \cite{aeroeff07}, \cite{weathereff}) and photometric calibration \cite{FDcalicrc07}.  

\begin{figure}[h]
\begin{minipage}{18pc}
\includegraphics[width=18pc]{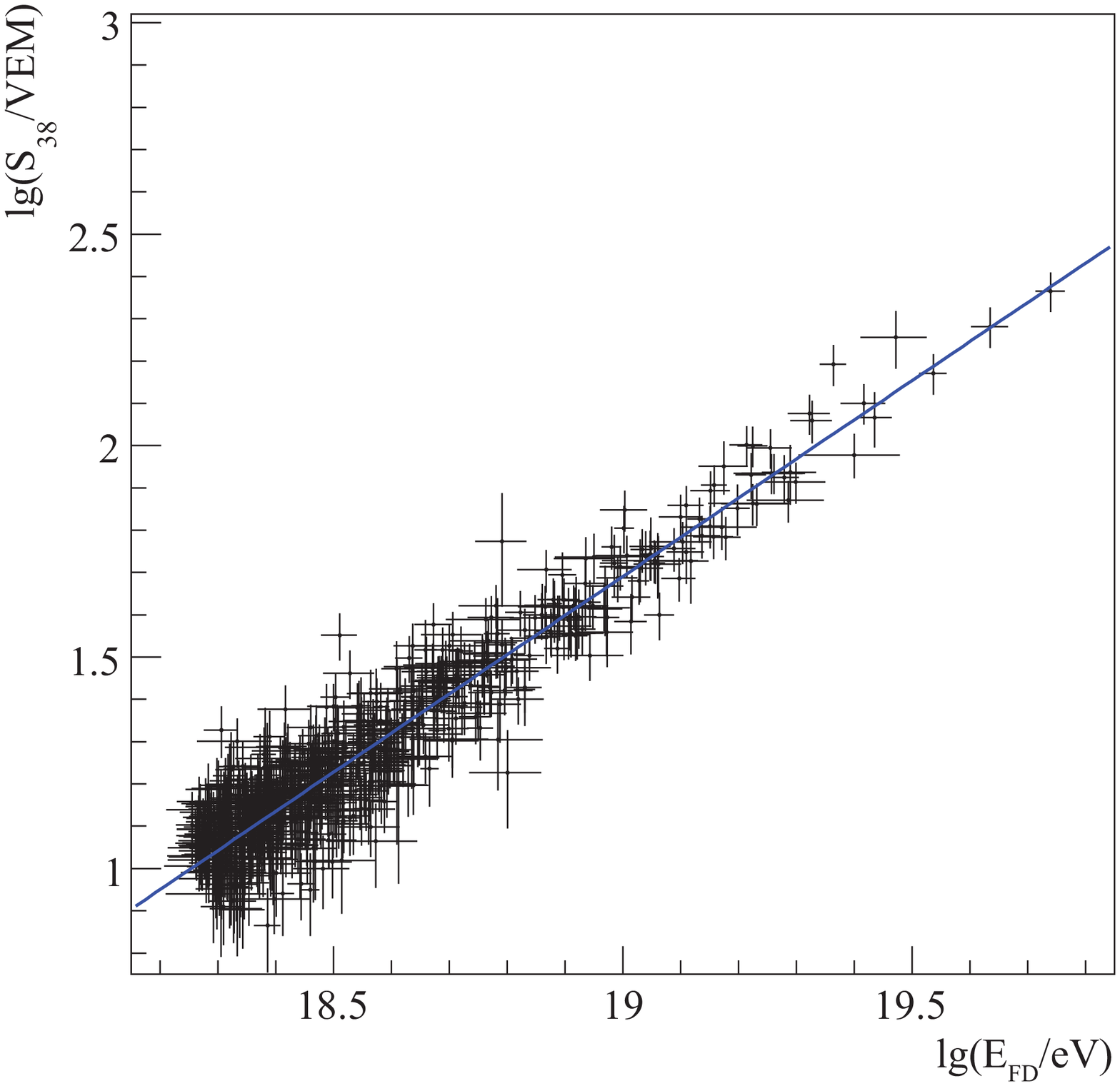}
\caption{\label{calib-scatter}Correlation between energy as determined by the FD (horizontal axis) and by the SD for selected hybrid cosmic ray events.}
\end{minipage}\hspace{2pc}%
\begin{minipage}{18pc}
\includegraphics[width=18pc]{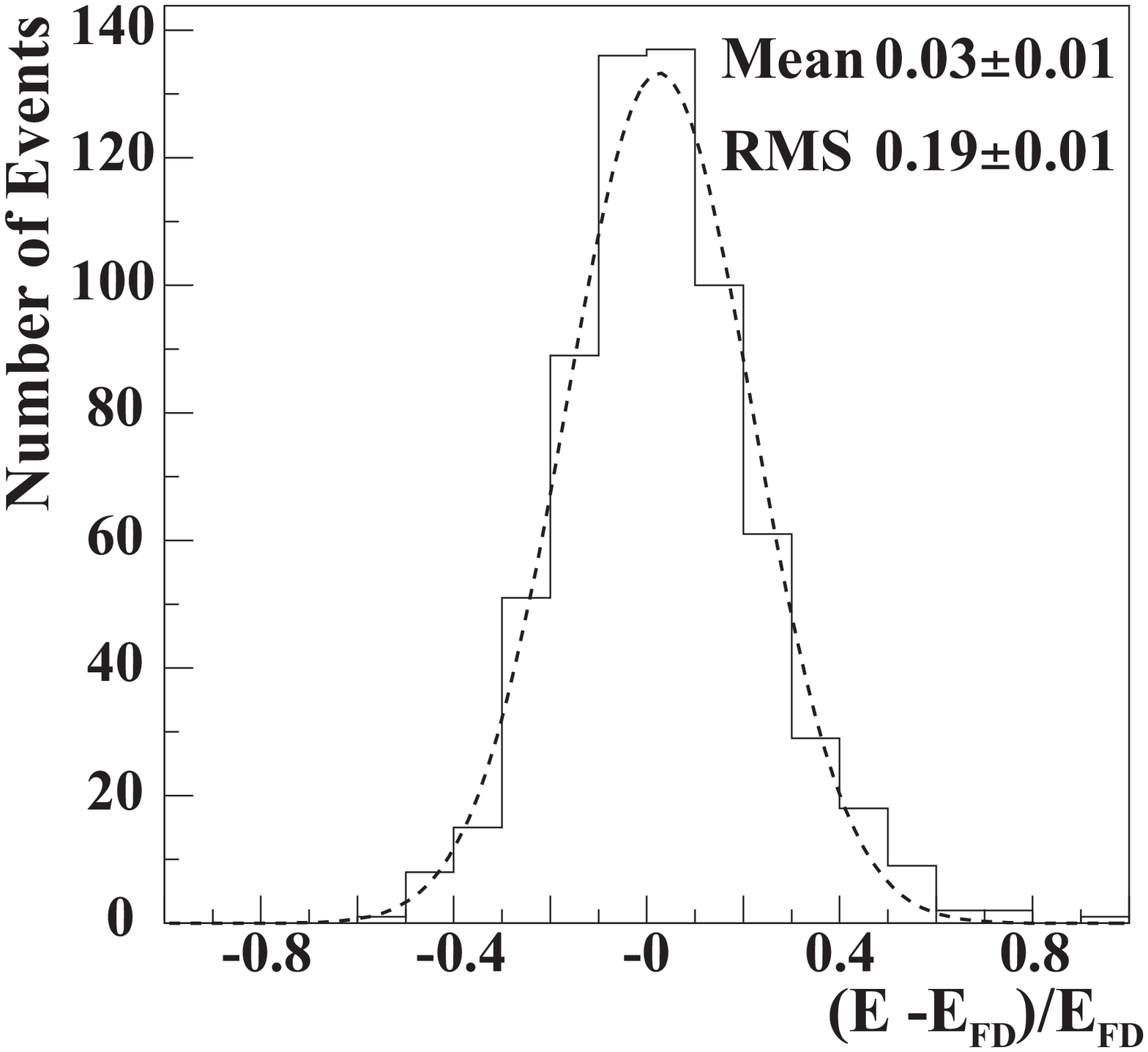}
\caption{\label{calib-histo}Energy correlation. E is the SD energy as calibrated by the FD using the fit in figure \ref{calib-scatter}.}
\end{minipage} 
\end{figure}

The SD operates 24/7 to provide the statistical engine for Auger South. It is nearly fully efficient at $5\times10^{18}$ eV. After correcting for zenith angle, the primary particle energy is estimated from the density of charged particles 1 km from the EAS core. Angular accuracy is a function of the number of tanks triggered. Essentially all events above the design threshold of $10^{19}eV$ trigger 5 or more tanks. For EASs measured by at least 5 tanks, this accuracy is better than 1 degree \cite{CB2005}.

The (FD) operates at night and records the longitudinal light profiles of EASs as
they develop through the atmosphere. This calorimetric energy measurement does not rely on interaction models; the amount of fluorescence light emitted is essentially proportional to the energy deposited in the atmosphere. 

About 10\% of the observed EASs are observed by the FD and by the SD. This hybrid data sample currently exceeds 100,000 events.  A subset, dubbed \textgravedbl golden hybrids\textacutedbl , can be reconstructed using either the FD or the SD. FD energy measurements of 661 golden hybrids have been used to establish the energy scale (figure \ref{calib-scatter}) of the high energy cosmic-ray spectrum measured by the SD \cite{esprl}. The sigma of the correlation histogram (figure \ref{calib-histo}) is 19\%. In terms of particle physics calorimetry, the level of consistency may seem suprising good; the SD is a sparse single-layer calorimeter. However, the large numbers of particles ($\geq10^{9}$) comprising the typical EAS reduce measurement fluctuations. At present the systematic energy uncertainty is less than 25\%. The three largest terms are fluorescence yield, detector photometric calibration, and atmospherics.  Laser test beams play a critical role in the last two terms. 

A motivation for reducing these systematics further is the strong dependence on energy of the propagation distance for protons through the CMBR. In the vicinity of the GZK suppression, a  $25\%$ decrease in energy scale from $8\times10^{19}$ eV to $6\times10^{19}$ eV corresponds to a 10-fold increase of observable volume.

\section{Laser "Test Beams"}
A variety of pulsed ultraviolet lasers (see table \ref{lasertable}) are fired routinely
into the atmosphere above the observatory. Their locations are shown in the map of figure \ref{config}.

\begin{table}[h]
\caption{\label{lasertable}Lasers used at the Pierre Auger Observatory.}
\begin{center}
\begin{tabular}{lllllll}
\br
System&Type&Wavelength&Max. Energy& Width&Direction & Receiver\\
\mr
Roving & Nitrogen & 337.1 nm&100 $\mu J$ & 4 ns & vertical& FD\\
CLF & YAG & 355 nm & 7 mJ & 4 ns & vert.+steered& FD\\
XLF & YAG & 355 nm & 7 mJ & 7 ns & vert.+steered& FD\\
LIDARs(4) & YLF & 351 nm & 100 $\mu J$ & 25 ns & steered& dedicated\\
\br
\end{tabular}
\end{center}
\end{table}

When a pulsed UV laser is shot into the sky, the atmosphere scatters light out of the beam. The same FD detectors that record tracks produced by the passage of cosmic ray air-showers downward through the atmosphere can also record tracks produced by the passage of laser pulses upward through the atmosphere. But unlike cosmic-rays, the laser energy, absolute firing time (using GPS \cite{gpsy1}), and direction can be specified and the pulses can repeated as necessary.

\begin{figure}[h]
\includegraphics[width=26pc]{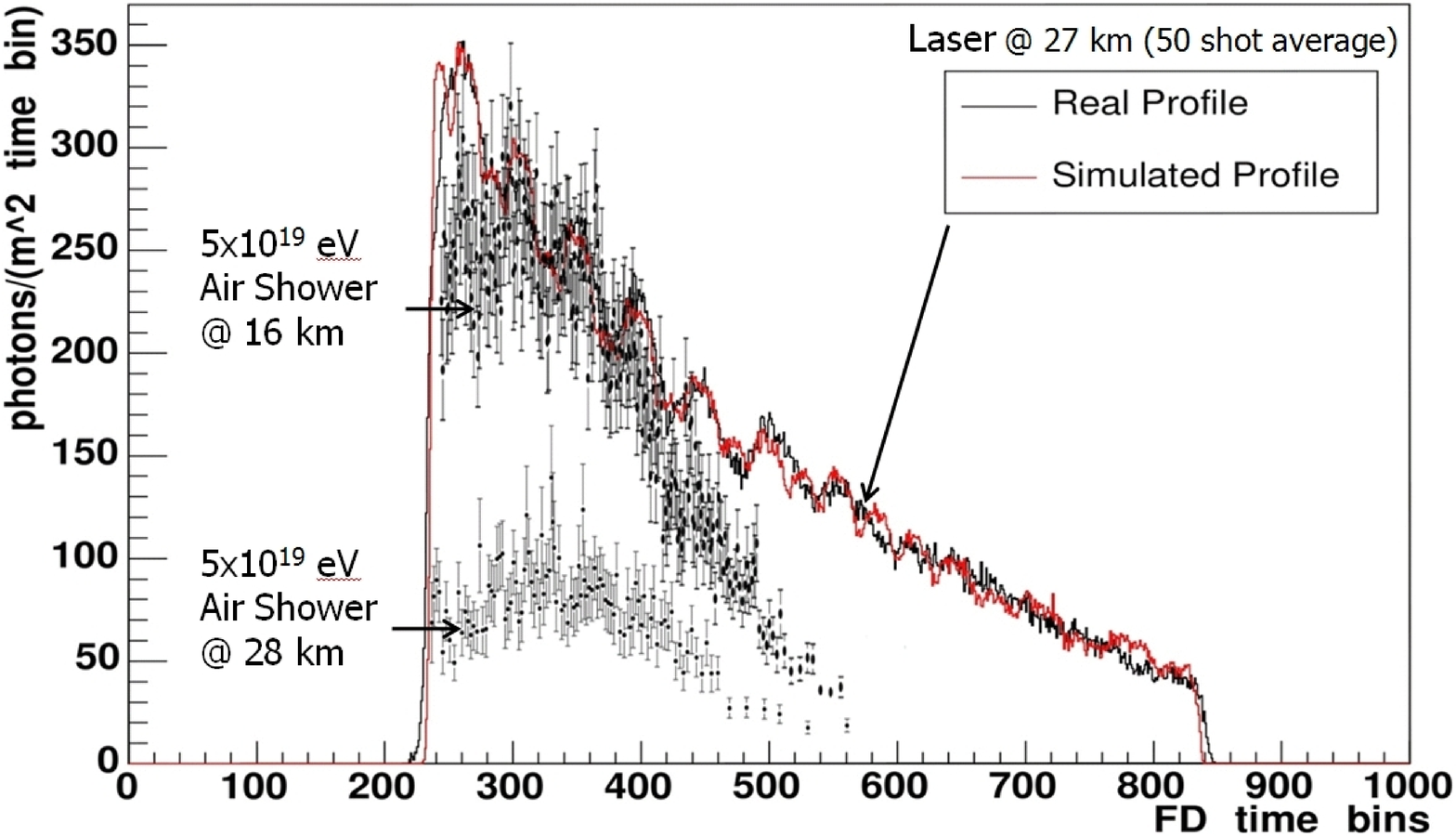}
\hspace {1pc}%
\begin{minipage}[b]{11pc} \caption{\label{l-se} Comparison between vertical 7 mJ UV laser shots from the CLF and near-vertical cosmic ray air-showers observed by the FD. One time bin is 100 ns. The superimposed (downward going) cosmic ray profiles have been \textgravedbl flipped\textacutedbl \space so that the left edge of all profiles corresponds to the bottom of the FD field of view.}
\end{minipage}
\end{figure}

Superimposed profiles from vertical laser tracks and near vertical (downward) cosmic-rays are shown in figure \ref{l-se} to illustrate the equivalence between laser energy and air-shower energy. Under similar conditions of atmospheric clarity, a vertical laser fired from the central laser facility generates a track in the FD that is about twice as bright as a track from a near-vertical cosmic ray of $5\times10^{19}$ eV seen at a similar distance of ~28 km. The observed laser and cosmic ray optical profiles in this example are similar in that the brightest regions of both are near the bottom of the FD field of view. Their shapes differ because different interaction processes are involved. The laser profile is generated by atmosphere scattering which, in this example, largely traces the decrease in atmospheric density with height. By contrast, the cosmic ray profile viewed from the side is dominated by scintillation light emission along the electromagnetic component of the air-shower. However, once the light from either the laser beam or the air-shower leaves the interaction region, both its propagation through the atmosphere and its conversion to digitized traces in the FD follow the same process.

\subsection{Roving Nitrogen Laser}
\begin{figure}[b]
\begin{minipage}{14pc}
\includegraphics[width=14pc]{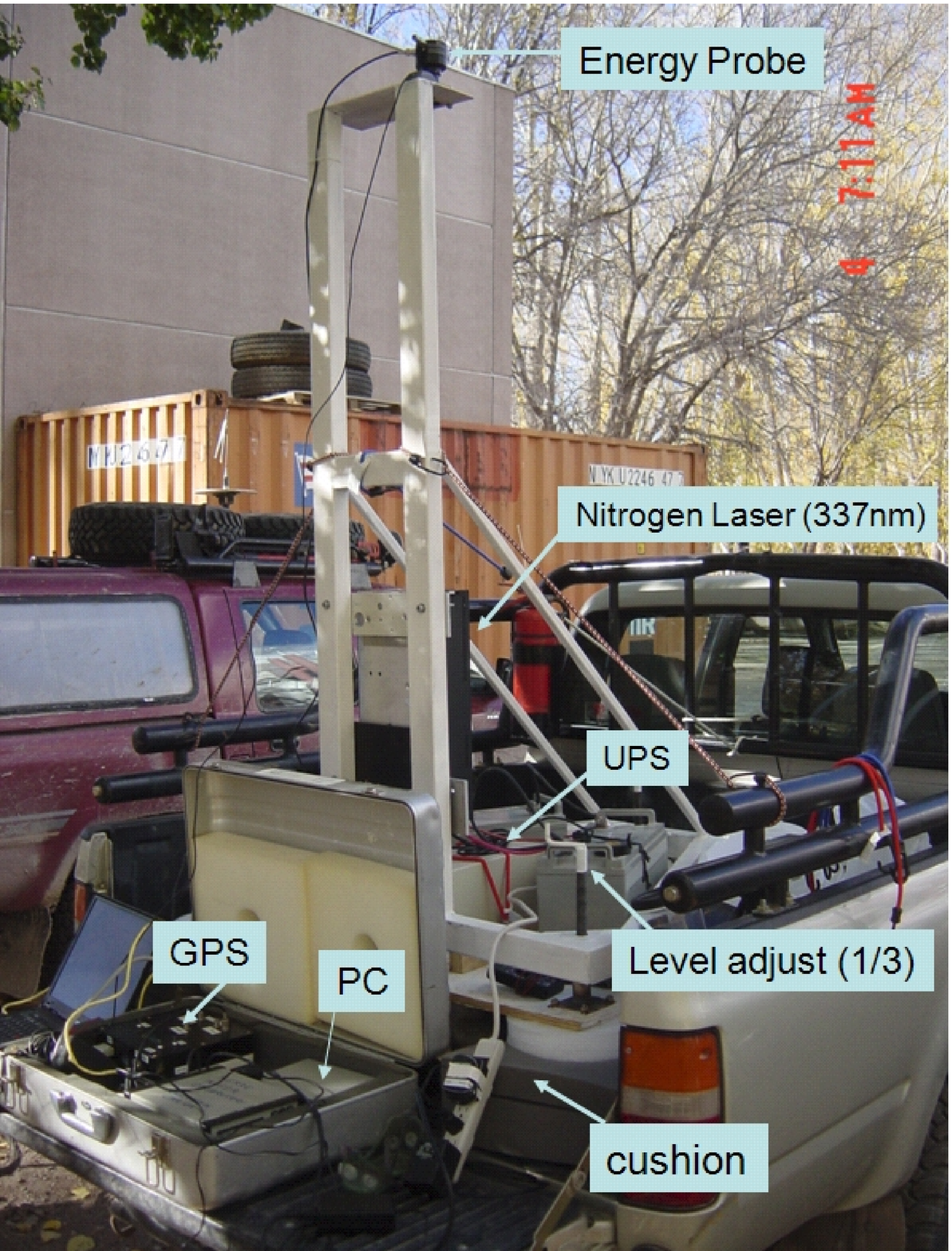}
\caption{\label{rov-pic}The roving nitrogen laser system is mounted on a small truck  for deployment in the field.}
\end{minipage}\hspace{3pc}%
\begin{minipage}{20pc}
\includegraphics[width=20pc]{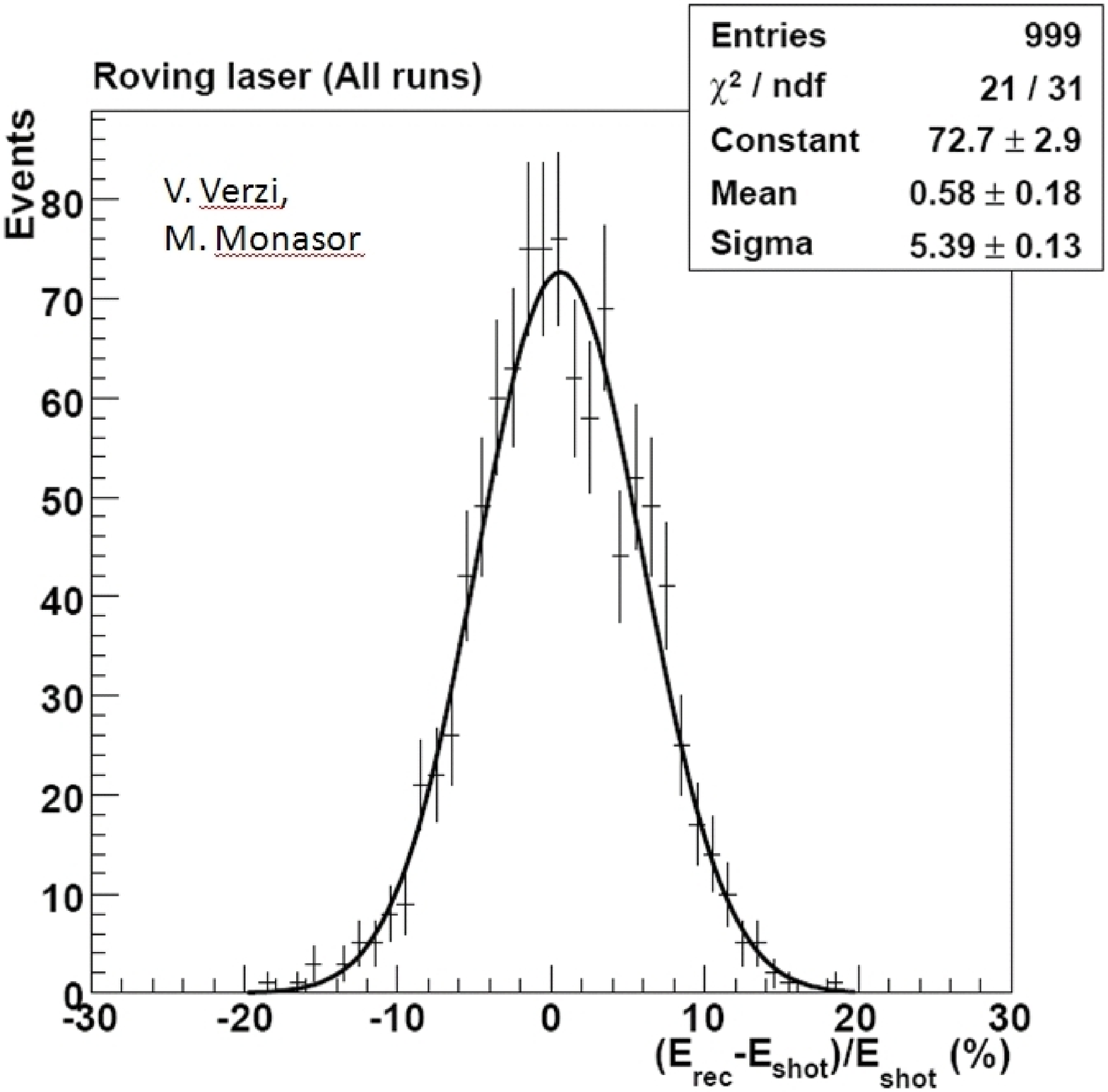}\hspace{2pc}%
\caption{\label{rov-eng} Cross-check of FD photometric calibration with the roving nitrogen laser system.  $E_{rec}$ is the energy reconstructed using the FD. $E_{shot}$ is the energy measured by a probe at the laser.}
\end{minipage} 
\end{figure}

A calibrated portable nitrogen laser system was configured (figure \ref{rov-pic}) to cross-check the photometric calibration of the FD. To reduce atmospheric effects the laser was fired vertically at distances of ~3km from the FD eyes. At this distance, the atmospheric effects of scattering out of the beam (that would brighten the track) and subsequent attenuation from the laser beam to the detector (that would dim the track) tend to cancel \cite{rflash}. The laser energy was measured by a calibrated energy probe and by reconstructing the tracks in the FD after the FD was calibrated using a "drum" light source placed over the telescope aperture. These independent energy measurements (see figure \ref{rov-eng}) were consistent with the 7\% accuracy of the energy probe and the 10\% accuracy of the FD photometric calibration.

\begin{figure}[t]
\includegraphics[width=21pc]{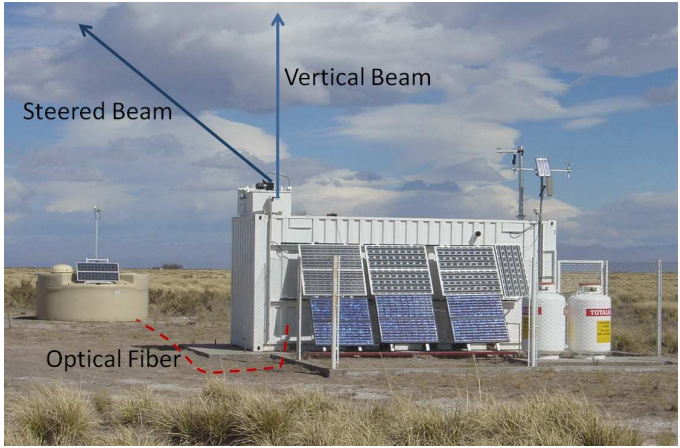}
\hspace {2pc}%
\begin{minipage}[b]{12pc}\caption{The Central Laser Facility at the Pierre Auger Southern Detector generates either a fixed-direction vertical or steered beam. A small fraction of light from each laser pulse is sent via optical fiber to the adjacent surface detector station. The facility is solar powered and is operated remotely from the central campus in Malargue.}
\end{minipage}
\end{figure}

\subsection{Central Laser Facility (CLF) and Extreme Laser Facility (XLF)}
\begin{figure}[h]
\begin{minipage}{18pc}
\includegraphics[width=18pc]{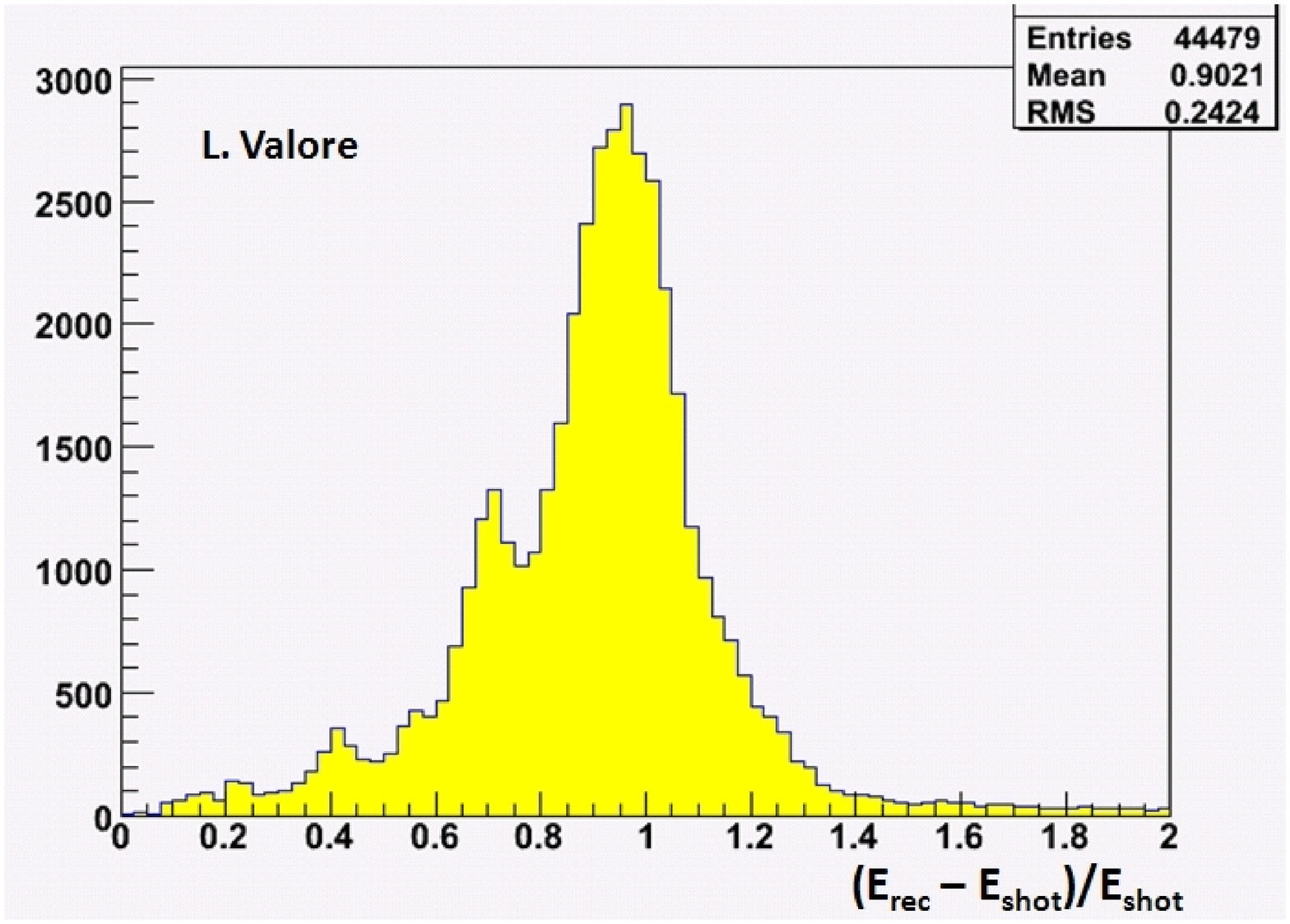}
\caption{\label{clf-rmol}Reconstructed CLF energy using tracks observed in the Los Leones FD eye 27 km distant. This reconstruction assumed a clear atmosphere. Aerosol and cloud effects broaden the distribution.}
\end{minipage}\hspace{2pc}%
\begin{minipage}{18pc}
\includegraphics[width=17pc]{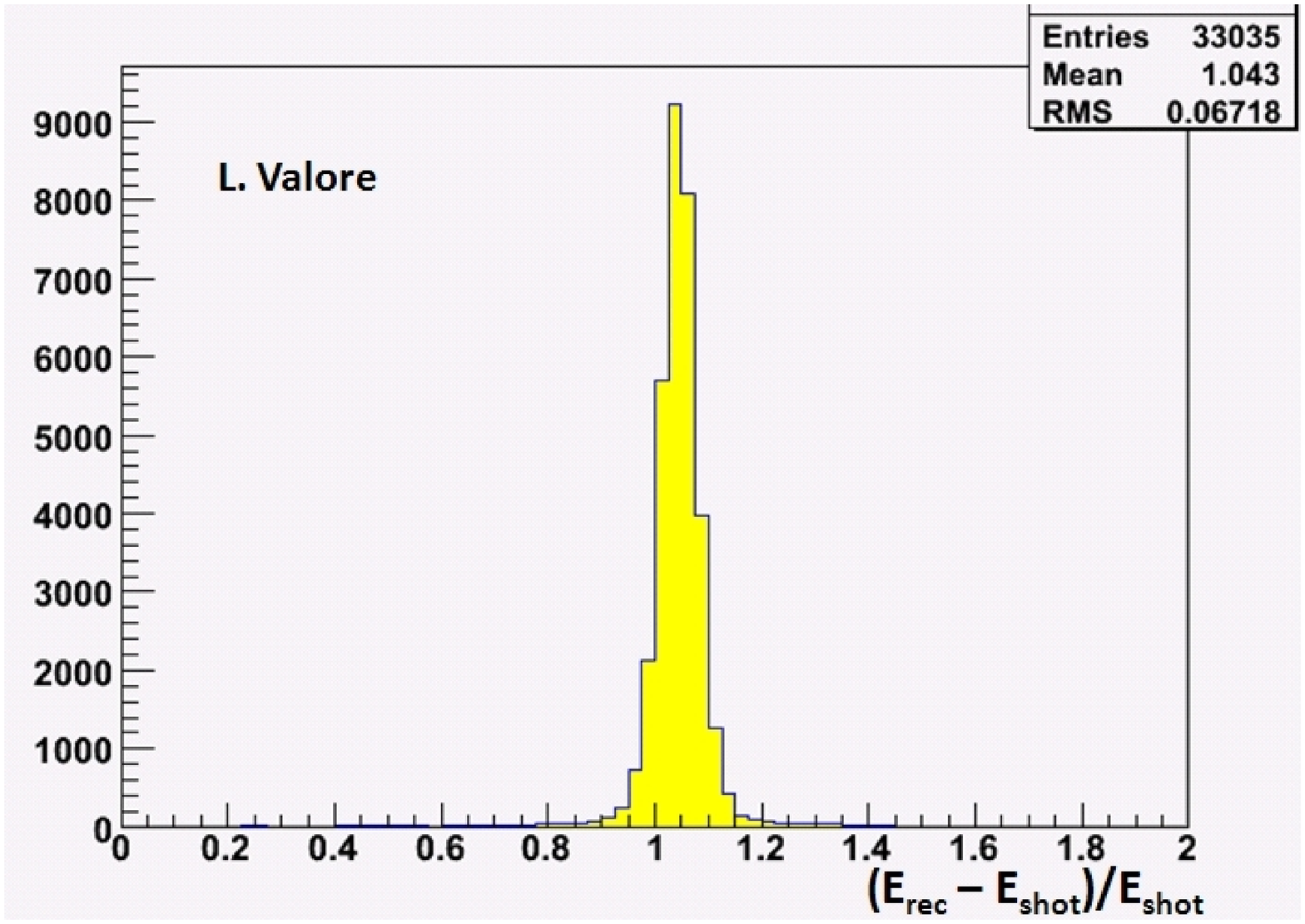}
\caption{\label{clf-ratm}CLF laser data in previous figure, but reconstructed with the hourly atmospheric database corrections. Cloudy periods are also removed. Since the atmospheric  corrections were obtained using CLF data, this cross-check is not robust.}
\end{minipage} 
\end{figure}

Laser test beams from the CLF have been used extensively at the observatory \cite{xpisa}. Details of the CLF can be found in \cite{clfjinst}. Its uses include FD commissioning (swapped cable finding), measuring and monitoring the offsets between the FD and SD  clocks, testing the FD geometric reconstruction, hourly monitoring of the atmospheric aerosol vertical distribution and horizontal uniformity, and detection of clouds in the center of the observatory. The CLF will also supplement the LIDAR systems in a \textgravedbl shoot-the-shower\textacutedbl \space program.  A preliminary cross-check of the photometric calibration of the FD has been performed with the CLF vertical beam (see figures \ref{clf-rmol} and \ref{clf-ratm}). This cross-check includes the effect of atmospheric attenuation over the 27-30 km distance from the CLF to the FD eyes. At this point this cross-check is not robust because the atmospheric corrections applied were those determined by the CLF.

\begin{figure}[t]
\centering
\includegraphics[width=24pc]{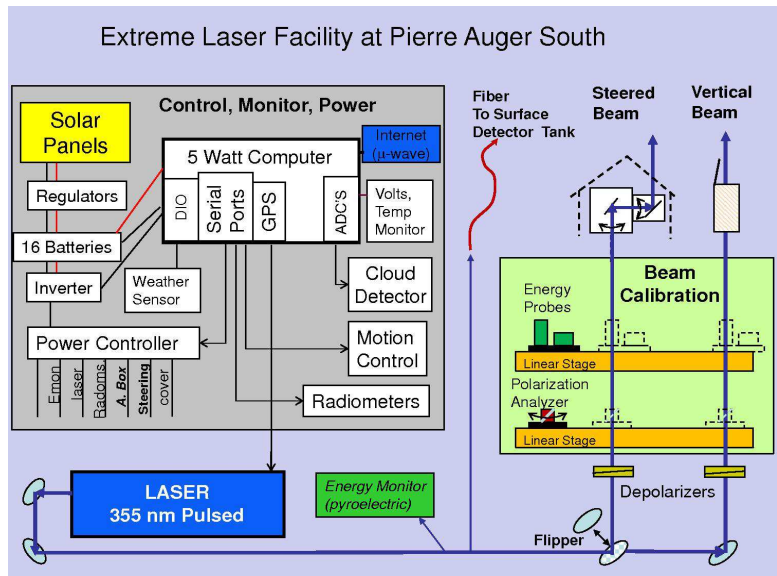}\hspace{2pc}%
\caption{\label{l-s}Functional diagram of the XLF  The beam energy and polarization calibration will be measured remotely.}
\end{figure}

A second laser facility, dubbed the XLF, will provide an independent cross-check of FD photometric calibration including atmospheric effects. The XLF design is similar to that of the CLF with the addition of an automated system to measure the absolute beam energy and polarization each night of operation (see figure \ref{l-s}). The goal is to maintain the absolute energy calibration of the beam including polarization effects to an uncertainty of less than 5\%.

\section{Auger North}
The highest energy events recorded by Auger South are not distributed isotropically in the sky \cite{science07}. However, despite the large collecting area, the statistics from the south are limited to some 30 events per year above $10^{19.5}$ eV. To collect a sufficent sample to identify and study sources, a companion Auger North detector planned for Colorado \cite{north} will be much larger than Auger South. The design features 4400 tanks covering 21,000 $km^{2}$ with $\sqrt{2}$ mile spacing. An FD of 4-6 stations will overlook most of the SD.  An extensive R\&D program is underway.

\end{document}